\documentclass[12pt, oneside]{article}   	
\usepackage{geometry} 
\usepackage[utf8]{inputenc}               		
\usepackage{graphicx}
\usepackage{epstopdf}
\usepackage{graphicx}				
\usepackage{amsmath}								
\usepackage{amssymb}
\usepackage{color}
\usepackage[usenames,dvipsnames,svgnames,table]{xcolor}
\usepackage[colorlinks=true,
            linkcolor=red,
            urlcolor=magenta,
            citecolor=blue]{hyperref}

\title{Wave-packet spreading in disordered soft architected structures }

\author{\fontsize{10}{10}\selectfont Arnold Ngapasare$^{1,2}$, \ Georgios Theocharis$^{2}$, \ Olivier Richoux$^{2}$\\
\fontsize{10}{10}\selectfont \ Charalampos Skokos$^{1}$  \ and \  Vassos Achilleos$^{2}$\\
\fontsize{10}{10}\selectfont $^{1}${Nonlinear Dynamics and Chaos Group} \\
\fontsize{10}{10}\selectfont {Department of Mathematics and Applied Mathematics} \\  
\fontsize{10}{10}\selectfont {University of Cape Town, Rondebosch 7701, South Africa. } \\  
\fontsize{10}{10}\selectfont $^{2}${ Laboratoire d'Acoustique de l'Universit\'{e} du Mans (LAUM)}\\
\fontsize{10}{10}\selectfont { UMR 6613, Institut d'Acoustique - Graduate School (IA-GS), CNRS, Le Mans Universit\'{e}, France }}   
\date{}							

\begin{document}
\maketitle

\begin{abstract}
{\fontsize{10}{10}\selectfont 
We study the dynamical and chaotic behavior of a disordered one-dimensional elastic mechanical lattice which supports translational and rotational waves. The model used in this work is motivated by the recent experimental results of B. Deng \textit{et al.,} Nat. Commun. $\mathbf{9}$, 3410 (2018). This lattice is characterized by strong geometrical nonlinearities and the coupling of two degrees-of-freedom (DoFs) per site. Although the linear limit of the structure consists of a linear Fermi-Pasta-Ulam-Tsingou lattice and a linear Klein-Gordon (KG) lattice whose DoFs are uncoupled, by using single site initial excitations on the rotational DoF, we evoke the nonlinear coupling between the system’s translational and rotational DoFs. Our results reveal that such coupling induces rich wave-packet spreading behavior in the presence of strong disorder. In the weakly nonlinear regime, we observe energy spreading only due to the coupling of the two DoFs (per site) which is in contrast to what is known for KG lattices with a single DoF per lattice site, where the spreading occurs due to chaoticity. Additionally, for strong nonlinearities, we show that initially localized wave-packets attain near ballistic behavior in contrast to other known models. We also reveal persistent chaos during energy spreading, although its strength decreases in time as quantified by the evolution of the system’s finite-time maximum Lyapunov exponent. Our results show that flexible, disordered and strongly nonlinear lattices are a viable platform to study energy transport in combination with multiple DoFs (per site), also present an alternative way to control energy spreading in heterogeneous media.}
\end{abstract}


\section{Introduction} \label{sec1}

Wave propagation in heterogeneous complex media has been a subject of intensive research interest in recent years. 
Among various systems, a large part of the conducted studies has been concentrated in families of one-dimensional ($1$D) continuous and discrete models~\cite{dyson,disorder_14,izrailev,disorder_15}
by focusing mainly on the localization properties of both the normal modes of finite systems i.e.,~Anderson localization (AL)~\cite{anderson}, as well as wave propagation in infinite media. 
The successful extension of AL to many other systems after it was initially formulated for electronic systems, has opened many research frontiers and applications \cite{bruinsma,schwartz,lahini,billy,roati}. Experimental results on AL (see e.g.,~Refs.~\cite{schwartz,lahini,billy,roati,jk}) have stimulated further interest in AL for both quantum and classical systems. 
 
Regarding linear disordered $1$D lattices, among different systems, special attention has been given to the tight binding electron model \cite{harrison},
the linear Klein-Gordon (KG) lattice~\cite{rogers} and the harmonic lattice \cite{dyson,Ishii,Ishii_2,Kundu}. These models are not only relevant to various physical systems but also represent the linear limits of seminal nonlinear lattices such as the discrete nonlinear Schr\"{o}dinger equation (DNLS), the quartic KG, and the Fermi-Pasta-Ulam-Tsingou (FPUT) lattices~\cite{fpu,kevre,izrailev,ggg}. Within the context of phononic and photonic lattices, these fundamental models have been adopted to describe a variety of physical systems and more recently, they have been used as a testbed for novel wave phenomena~\cite{deymer2,flachbook}.

A common route to study the wave properties of disordered lattices is through monitoring the time evolution of initially compact wave-packets. For tight binding and linear KG models, the dynamics after the excitation of such an initial condition is characterized by a transient initial phase of spreading, followed by a phase of total confinement to the system's localization length. The width of the wave-packet is of the order of the maximum localization length~\cite{loc_len}. On the other hand, for the harmonic lattice, along with the localized portion of the energy, there is always a propagating part due to the existence of extended modes at low frequencies. A quantitative description of wave propagation in disordered $1$D systems of one degree of freedom (DoF) per lattice site was formulated in Refs.~\cite{Ishii, Kundu, Lepri} where wave-packet spreading was quantified using both analytical and numerical methods. Moreover, many variations of these $1$D lattices have been studied extensively in several works including dynamical regimes ranging from the homogeneous linear to the disordered nonlinear~\cite{Chiaradis,Guebelle,MasonPanos,vassospre2016,Luding2,Sen2017poly,jk,arn_gran,many2}. 

As a natural extension to the above studies, an investigation into the corresponding behavior in disordered lattices with more than one DoF per site seems plausible. Not many studies along such lines have been reported in the literature. The majority of existing works have taken the approach of making generalizations of the tight binding model by assuming a linear coupling between two (or more) $1$D chains~\cite{ladderPRB,flachlieb}. Such coupling results in changes to the dispersion relation thereby changing the energy transport properties. Our recent work with a linear disordered phononic lattice \cite{arn_phononic} is indeed an attempt to fill this gap. The wave dynamics of disordered harmonic chains with two DoFs per site appear to be quite interesting, deserving further investigations. Such systems have been useful in modeling macroscopic mechanical devices including granular chains, highly deformable elastic assemblies and origami lattices~\cite{pichard,florian1,yasuda,yasuda2,deng2,florian2}. This allows for easy  tunability of the system's dispersion due to the geometrical characteristics and material properties, and makes these systems attractive for several applications.
 
Here we focus on highly deformable architected lattices, characterized by a nonlinear response which enabled the design of new classes of tunable and responsive elastic materials. Several such soft structures have already been reported including bioinspired  soft robots \cite{shepherd, yang}, self-regulating microfluidics \cite{Beebe}, reusable energy absorbing systems \cite{shan,restrepo}, materials with programmable response \cite{florijn} and information processing via physical soft
bodies \cite{nakajima}. Furthermore, soft architected materials present opportunities to control the propagation of elastic waves, since their dispersion properties can be altered by applying a large, nonlinear pre-deformation \cite{boyce1,boyce2,casadei} or changing the geometry \cite{deng2}. To date, most of the investigations have predominantly focused on linear stress waves, or soliton solutions of such systems due to the capability of the soft structures to support large-amplitude nonlinear waves. Here, taking a step forward, we study a particular lattice  that supports both translational and rotational waves~\cite{deng2}. Our main goal is to understand how nonlinear lattice waves propagate in the presence of strong disorder when the DoFs are coupled, as well as the system's chaoticity.
 
The rest of this paper is arranged as follows: In Section~\ref{sec2} we describe the Hamiltonian model of the lattice structure and also formulate the system's equations of motion. The dispersion relation of the system, in addition to its dynamics in the linear limit is also discussed. In Section~\ref{sec3} we investigate the nonlinear effects on wave propagation under strong disorder, as well as study the system's chaoticity and finally, in Sec.~\ref{sec4}, we summarize our findings and present our conclusions.
\section{The Hamiltonian model and its equations of motion} \label{sec2}

\begin{figure}[t!] 
\centering
\includegraphics[width=12.20cm]{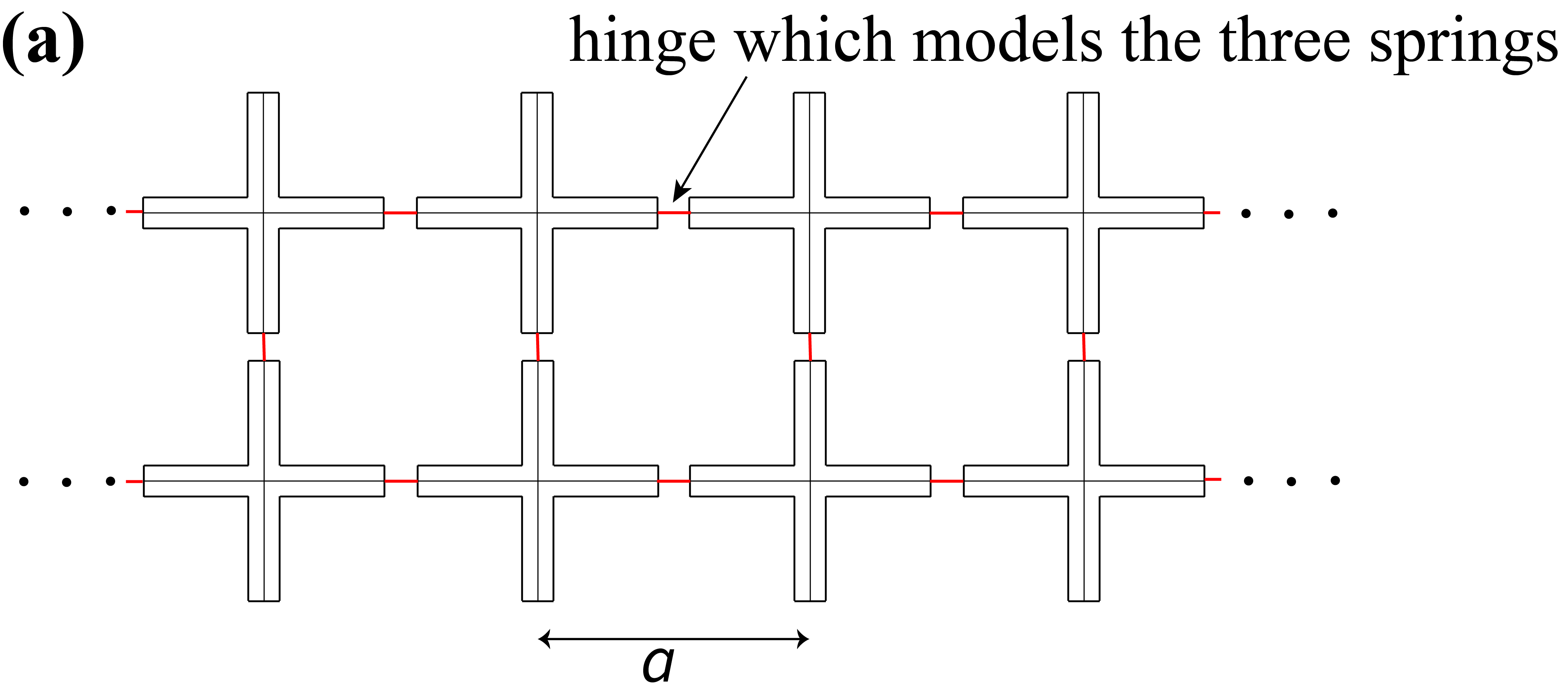} \\
\includegraphics[width=12.20cm]{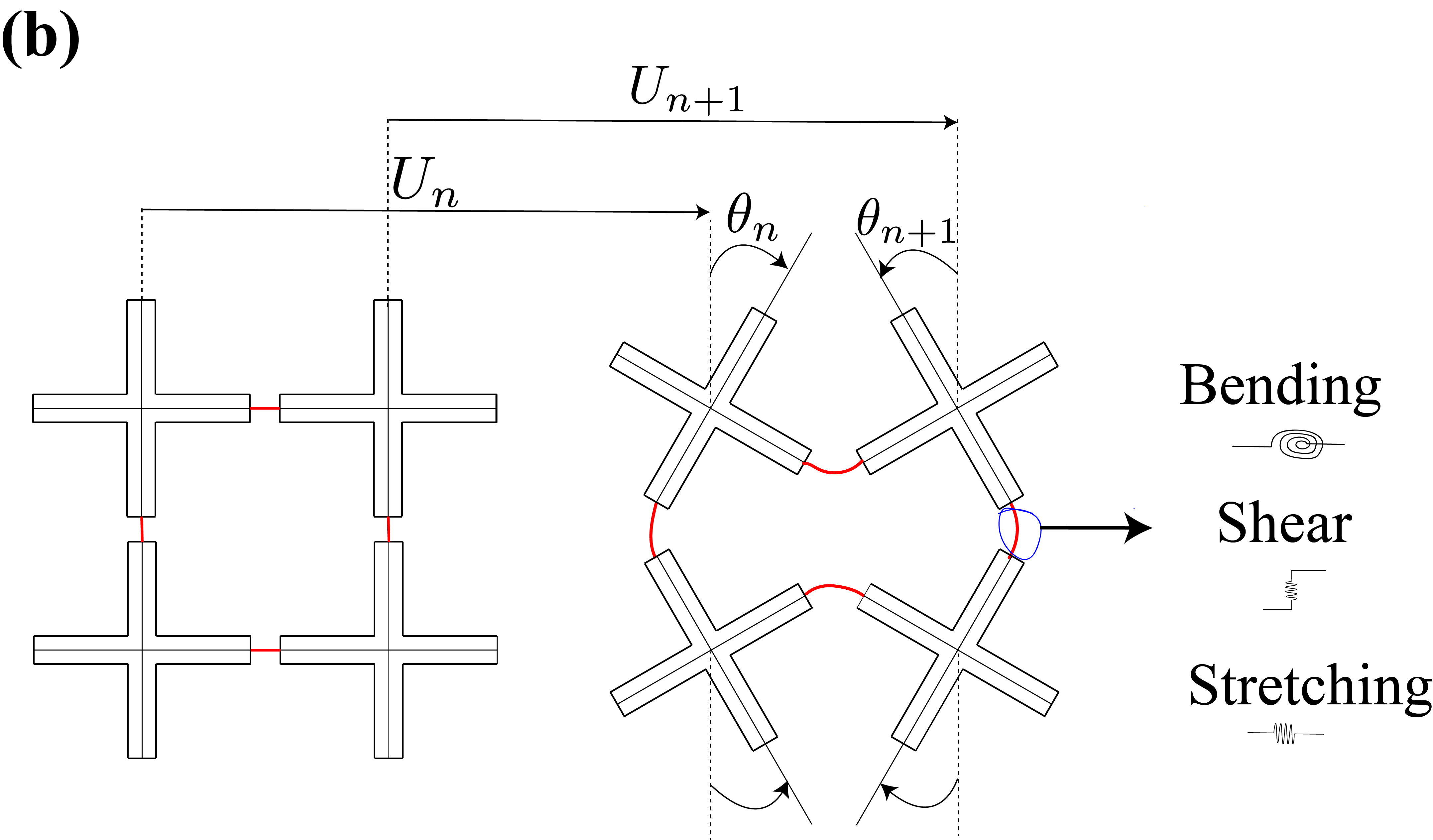}
\caption{\label{fig1} (a) An architected, highly deformable and elastic mechanical structure which supports translational and rotational waves. (b) Schematic of the cross pairs [by symmetry, the dynamics can be described by either the top row or bottom row of (a)] showing the translational and angular deflections deflections. The connectors (marked in red) model a combination of bending $k_{\theta}$, shear $k_s$ and stretching springs $k_l$.}
\end{figure}

The $1$D elastic mechanical lattice studied in this work is assembled from an array of aligned LEGO\textsuperscript{\textregistered} crosses connected by flexible links \cite{celli} as depicted in Fig.~\ref{fig1}. This system constitutes a highly deformable elastic lattice supporting both translational and rotational waves ($2$ DoFs per site). In Ref.~\cite{deng2}, the authors describe the general equations of motion for a structure that takes into consideration 
some of the possible geometrical variations of the lattice. However, for the purposes of this work, we limit ourselves to an aligned, symmetrical structure. The crosses are joined to their neighbors by some flexible hinges which are modeled using a combination of three linear springs. The stretching is modeled by a spring with stiffness $ k_{l}$ 
and the shearing is described by a spring $ k_{s}$, whilst the bending is modeled by a torsional spring $ k_{\theta}$ [see Fig.~\ref{fig1} (b)].

By making use of the spatial periodicity $a$, we recast the horizontal deflections $u_n$ at the $n$th lattice site to $U_n = u_n/a$ and change time units to dimensionless time, $ T = t\sqrt{ k_{l}/\overline{m} } $ and the springs are normalized as $K^{(\theta)} = 4 k_{ \theta }/k_{l} a^2$ and $K^{(s)} = k_{s}/k_{l}$. The mass of each unit of the cross $m_n$ is normalized such that $M_n = m_n/\overline{m}$ where $\overline{m} $ is the average mass of the crosses. This implies that in the case of the homogeneous chain, $M=M_n  = m_n/\overline{m} = 1$ and $\Gamma=\Gamma_n  = J_n/ \overline{m} a^2$. Herein, $J_n$ denotes the rotational moment of inertia at the $n$th lattice site.
For the rest of this work, we assume the parameters of the experimental setup of Ref.~\cite{deng2} where each cross has mass $m = 4.52$~g ($\overline{m}=m$), $a = 42$~mm and $J = 605$ g$\cdot$mm$^2$.
Also the spring constants are $k_{l}=71.69$ N$\cdot$mm$^{-1}$, $ k_{s} = 1.325 $N$\cdot$mm$^{-1}$ and $k_{\theta} = 4.85$ N$\cdot$mm.

The Hamiltonian $H$, of the top or bottom layer of the system is thus given as (see Ref.~\cite{deng2} for details)
\begin{equation}
H  =\sum_{n=1}^{N} \left\{  \frac{ M_n \dot{U}^2_n}{2} +   \frac{ \Gamma_n \dot{\theta}^2_n }{2} +  \frac{1}{2}  \Delta^{n~^{2}}_{_{LH}} + \frac{K^{(s)}}{2}  \Delta^{n~^{2}}_{_{SH}}   + \frac{K^{(\theta)} }{8 } \left(  \delta^{n~^{2}}_{_{\theta H}}  + \frac{1}{2}  \delta^{n~^{2}}_{_{\theta V}} \right) \right\},
\label{hamil} 
\end{equation}
where the dimensionless deflections are given by, \\
$ \Delta^{n}_{_{LH}}  = U_{n+1} - U_{n} + \frac{1}{2} \left( 2 - \cos{ \theta_{n}}  -  \cos{ \theta_{n+1}} \right),$
$ \Delta^{n}_{_{SH}}  = \frac{1}{2} \left( \sin{ \theta_{n+1}}  -  \sin{ \theta_{n}} \right),$
$  \delta^{n}_{_{\theta H}} = \theta_{n+1} + \theta_{n}, $ and   $\delta^{n}_{_{\theta V}} = 2 \theta_{n}.$
In Eq.~(\ref{hamil}), $[~\dot{}~]$ denotes the derivative with respect to time.

We derive the equations of motion from the Hamiltonian Eq.~(\ref{hamil}) which yields
\begin{equation}
\begin{aligned}
 M_n  \ddot{U}_{n}  & =  \Big[ U_{n+1} - U_{n} + \frac {1}{2} \{  2 -  \cos( \theta_n)  -  \cos( \theta_{n+1}) \}  \Big] \\
   - &\Big[ U_{n} - U_{n-1} + \frac {1}{2} \{ 2 -  \cos(\theta_n)  -  \cos( \theta_{n-1}) \}  \Big ],
\label{em1}
\end{aligned}
\end{equation}

 \begin{equation}
\begin{aligned}
 \Gamma_n \ddot{\theta}_{n} & =   \frac{1}{4} K^{(s)} \cos( \theta_n)  \Big[ \sin( \theta_{n+1})  -  \sin( \theta_{n})   \Big]  \\
  + &\frac{1}{4} K^{(s)}  \cos( \theta_n)  \Big[ \sin( \theta_{n-1})  -  \sin( \theta_{n})   \Big]  \\
+ &\frac{1}{4} \sin( \theta_n)  \Big[ 2 (U_{n} - U_{n+1}) + \cos( \theta_n)  + \cos( \theta_{n+1}) - 2  \Big] \\
 + &\frac{1}{4} \sin( \theta_n)  \Big[ 2 (U_{n-1} - U_{n}) + \cos( \theta_n)  + \cos( \theta_{n-1}) - 2  \Big]  \\
-&\frac{1}{4} K^{(\theta)} ( \theta_{n+1} + 4\theta_{n} + \theta_{n-1} ).
\label{em2}
\end{aligned}
\end{equation}

\subsection{  Homogeneous linear system} \label{ssec1}
First let us consider the homogeneous system 
and linearize the nonlinear terms (trigonometric terms) in 
Eqs.~(\ref{em1}) and (\ref{em2}) by assuming small angles, $\theta_{n+p}$ with $p = \left\{-1,0,1\right\} $,
and taking a power series expansion of the appropriate cosine and sine terms to give the first two lowest order terms as

\begin{equation}
 \sin \theta_{n+p} \approx  \theta_{n+p} - \frac{1}{6} \theta_{n+p} ^3 + \ldots,
\label{eq2}
\end{equation}

\begin{equation}
 \cos \theta_{n+p} \approx  1 - \frac{1}{2} \theta_{n+p} ^2 + \ldots.
\label{eq3}
\end{equation}
The linear parts of Eqs.~(\ref{eq2}) and (\ref{eq3}) are plugged appropriately into 
Eqs.~(\ref{em1}) and (\ref{em2}) to give the linear equations of motion as
\begin{equation}
\begin{aligned}
  M_n \ddot{U}_{n}  & =  U_{n+1} - 2 U_{n} + U_{n-1}  ,
\label{em3}
\end{aligned}
\end{equation}
and 
 \begin{equation}
\begin{aligned}
 \Gamma_n \ddot{\theta}_{n} & =  \tilde{K} (\theta_{n+1} -2\theta_{n} + \theta_{n-1}) - 6 K^{(\theta)} \theta_{n},
\label{em4}
\end{aligned}
\end{equation}
where $\tilde{K} = K^{(s)} - K^{(\theta)}$.
For the homogeneous case, $M_n = M = 1$ and $\Gamma_n = \Gamma$ as already defined. 
A quick glance at Eqs.~(\ref{em3}) and (\ref{em4}) reveals that the two sets of DoFs are decoupled in the linear regime. Eq.~(\ref{em3}) belongs to the linear FPUT class of equations and Eq.~(\ref{em4}) is a linear KG-type equation.

We now consider solutions of the form
\begin{equation}  
\mathbf{X}_n =  \begin{pmatrix} 
            U_n (t)\\
            \theta_n (t)
           \end{pmatrix} = \mathbf{X}e ^{i \omega t - i q n},
 \label{eq4}          
\end{equation}
 where $\mathbf{X}=[U_0,\Theta_0]$ is the amplitude vector, $\omega$ is the cyclic frequency
and $q$ is the wave number. Inserting Eq.~(\ref{eq4}) into Eqs.~(\ref{em3}) and ~(\ref{em4}), we obtain the following eigenvalue problem for the allowed frequencies 
$\mathbf{D} \mathbf{X} = \Omega^2 \mathbf{X}$,
where the resultant dynamical matrix is
$$ \mathbf{D} = \begin{pmatrix}
     2 -2 \cos q  & 0 \\
      0 &  \frac{1}{2 \Gamma}\left [(K^{(s)} + 2 K^{(\theta)}) - ( K^{(s)} - K^{(\theta)}) \cos q \right ]  
         \end{pmatrix}.
         $$
The corresponding dispersion branch for the transverse DoFs is
\begin{equation}
 \omega^{(U)} = \sqrt{2 -2 \cos q},
\label{eq5}
\end{equation}
whilst for the rotational DoFs,
\begin{equation}
 \omega^{(\theta)} = \frac{1}{\sqrt{4 \Gamma}} \sqrt{2\left (K^{(s)} + 2 K^{(\theta)}) - 2( K^{(s)} - K^{(\theta)}\right ) \cos q  }.
\label{eq6}
\end{equation}

\begin{figure} 
\centering
\includegraphics[width=12.50cm]{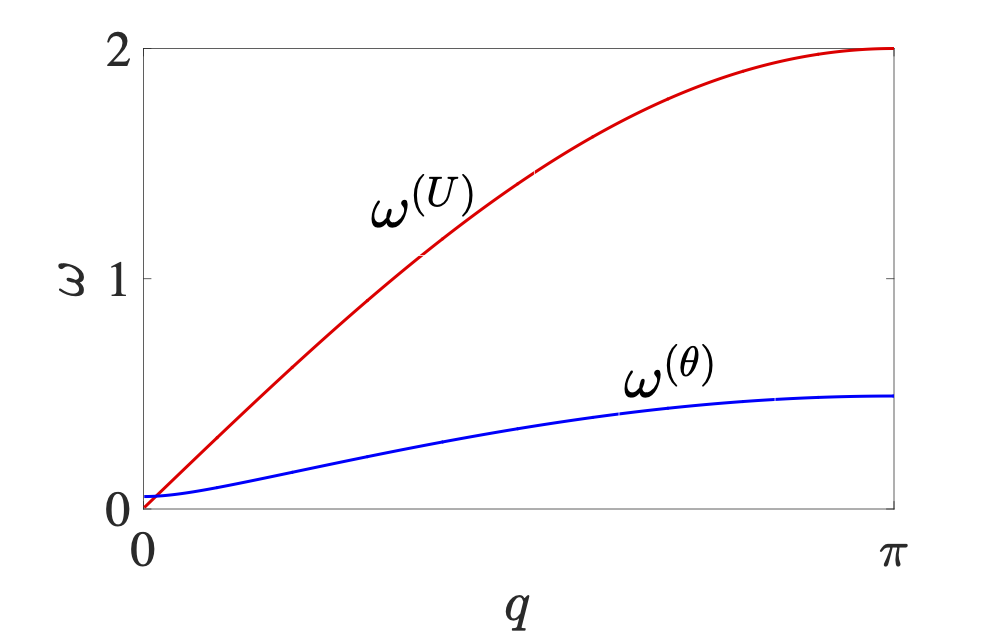}
\caption{\label{fig2} The two dispersion branches of the system for translational ($\omega^{(U)}$ - red curve) and rotational ($ \omega^{(\theta)}$ - blue curve) DoFs [See respectively Eqs.~(\ref{eq5}) and (\ref{eq6})].}
\end{figure}
The dispersion relation for the two different DoFs as given by Eqs.~(\ref{eq5}) and (\ref{eq6}), are plotted in Fig.~\ref{fig2}.
An important feature for the system is that the rotational mode branch (blue curve in Fig.~\ref{fig2}) starts at a finite frequency. This means that linear rotational waves are not supported for $\omega < \frac{1}{\sqrt{4 \Gamma}} \sqrt{6 K^{(\theta)}}$ since this frequency domain corresponds to a band gap. In fact, the linear dispersion relation of the rotations can be directly mapped onto the one for the linear KG lattice. 
On the other hand the transverse displacements follow a typical linear mass-spring dispersion relation.

\subsection{ Disordered linear system} \label{ss2b}
In order to consider to consider a disordered version of the system, we first note that the disordered linear KG system exhibits AL (See Ref.~\cite{Kundu} for the corresponding behavior of the linear FPUT). Additionally, even though there are a number of ways to introduce disorder in the system, we model the crosses of the architected lattice assuming disorder in the masses $M_n$, which in turn implies disorder in the rotational moments of inertia $\Gamma_n$. In practice, disorder in the system can also be achieved by changing the material used to manufacture the LEGO\textsuperscript{\textregistered} bricks at each site \cite{celli} without changing their geometrical dimensions. In dimensionless units, the masses are normalized to unity for a homogeneous chain hence we take this into consideration when choosing the disorder distribution and take $M_n$ from a uniform probability distribution $ f \big( M_n \big) $ where
\[
 f \big( M_n \big)  =
  \begin{cases}
                                   W^{-1}, &-W/2    < \text{ $ M_n $} - 1 < W/2, \\
                                   0 & \text{otherwise}. \\
    \end{cases}
\]
$ W$ denotes the distribution width and for this study, we choose $ W = 1.8$ hence $ 0.1 \leq M_n \leq 1.9 $. 

To study the dynamical behavior of the system, the equations of motion are integrated using the \texttt{ABA864} symplectic integrator \cite{blanes} which has been proved to be very efficient for the accurate integration of large Hamiltonian lattice models \cite{senyange2018,danieli2019}. 
This integration scheme allows for energy conservation of the total energy $H$, and keeps the relative energy error $\Delta H (T)  = \left|\dfrac { H (T) - H (0)  }{ H (0) }\right| <  10^{-5}$ 
when the integration time step is set to be $ \tau = 0.1$. In all our numerical simulations, we employ fixed boundary conditions i.e.,~$ U_{0} = U_{N+1} = 0$, $ \theta_{0} = \theta_{N+1} = 0$ ,  $ \dot{U}_{0} = \dot{U}_{N+1} = 0$ and $ \dot{\theta}_{0} = \dot{\theta}_{N+1} = 0$. Furthermore, the considered lattice size is large enough so that the energy does not reach the lattice boundaries. A typical numerical integration of the nonlinear system for $ T = 10^5$ requires a lattice size of at least $2 \times 10^5$ sites. 

\subsubsection{Translational degrees of freedom}
\begin{figure} 
\centering
\includegraphics[width=6.20cm]{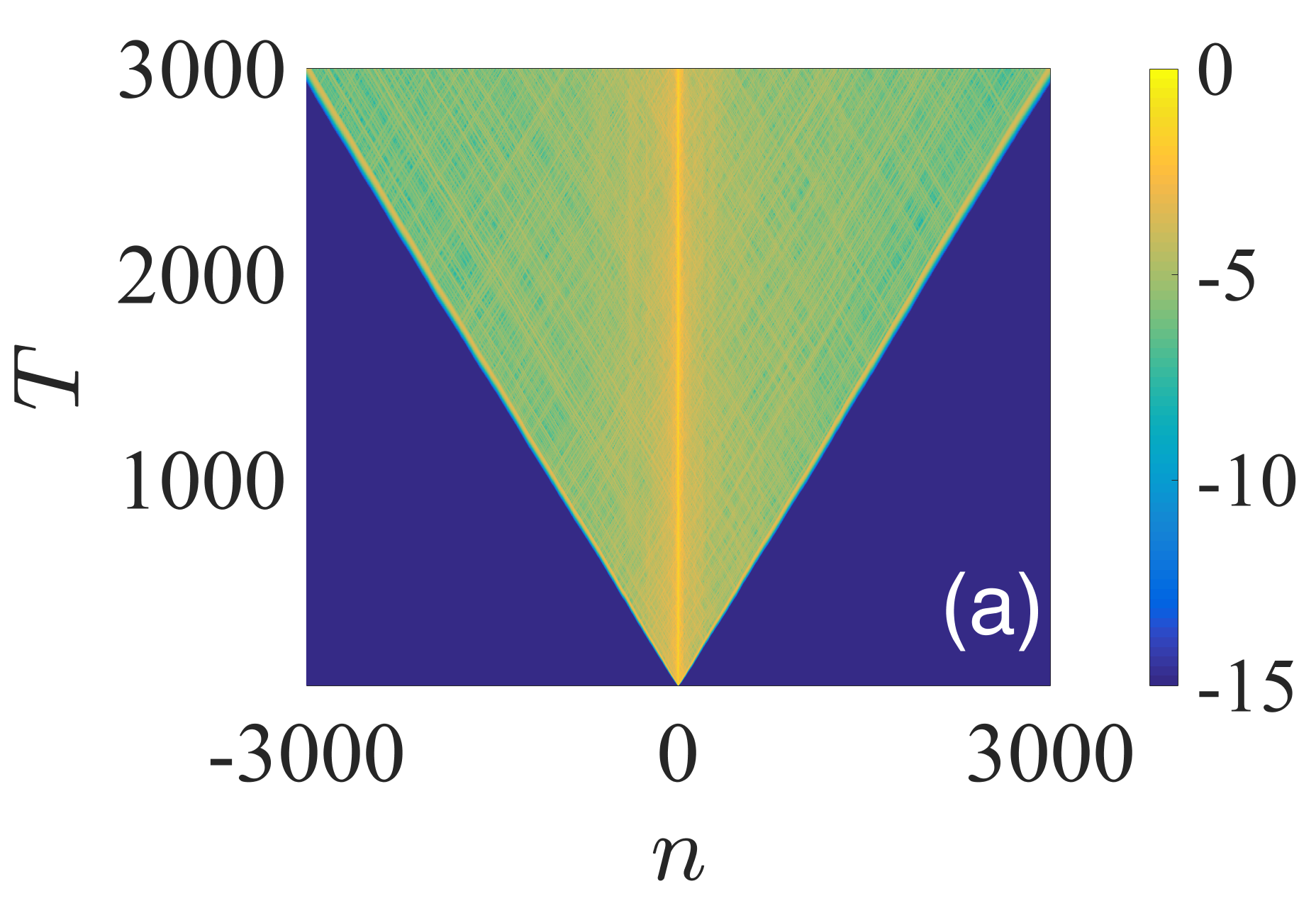}
\includegraphics[width=6.20cm]{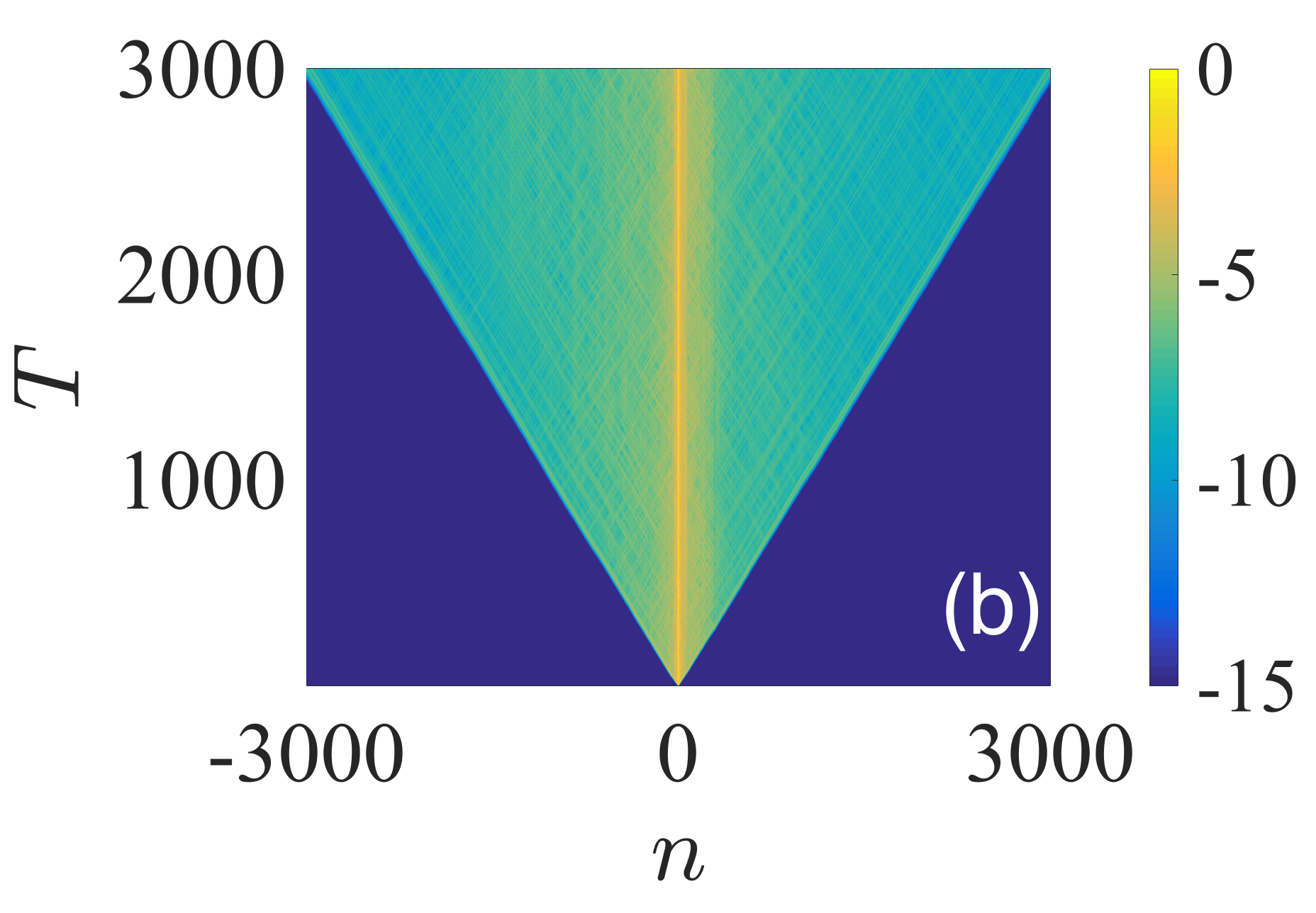}\\
\includegraphics[width=6.20cm]{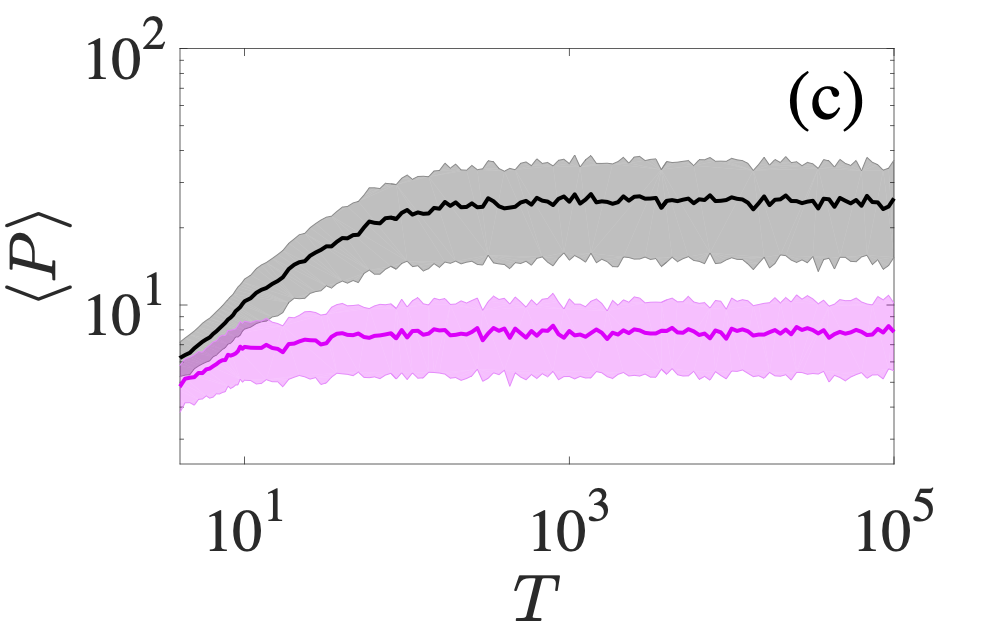}
\includegraphics[width=6.20cm]{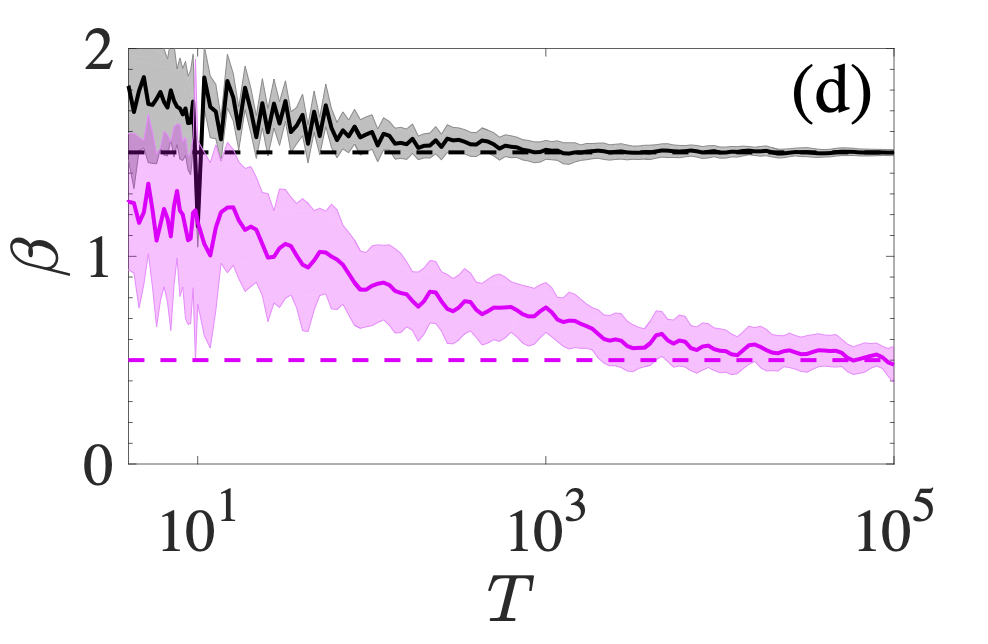}
\caption{\label{fig3} Spatiotemporal evolution of the energy distribution for a representative realization with (a) velocity and (b) displacement single site initial excitation in the linear system. The colorbars in (a) and (b) are in log-scale. (c) Time evolution of the average participation number $\langle P \rangle$, (d) estimation of the exponent $\beta$, related to the time evolution of the average second moment through $ \langle m_2 \rangle \propto T^{\beta} $. For velocity (black curves) and displacement (magenta curves) initial excitations, the mean values $ \langle \cdot \rangle$ are calculated from $100$ disorder realizations and the shaded areas represent the statistical error (one standard deviation). The dashed lines in (d) indicate $\beta =1.5$ (top) and $\beta =0.5$ (bottom) and the system energy for all realizations is $ H = 10^{-4} $. }
\end{figure}

We start our analysis by first exploring the two possible single site initial excitations of velocity and displacement for the translation DoFs i.e.,~

\begin{equation}
\dot{U}_{N/2}(0)=\sqrt{2H/M_{N/2}}, \quad \textrm{or} \quad U_{N/2}(0)= \nu,
\label{eq7} 
\end{equation}
independently. In Eq.~(\ref{eq7}), the scalar $\nu$ is real and its value is altered to match the desired system energy. We fix the total system energy at $H=H_{0}$ and integrate the system up to $T = 10^5$ time units and observe how the initially localized wave-packet evolves in time. A typical scenario of the dynamics is shown in Figs.~\ref{fig3}(a)-(b) for $H_0 = 10^{-4}$. For single site velocity initial excitations, the energy distribution displays two main parts, a central localized part, and an expanding peripheral part which is spreading beyond the excitation point [see Fig.~\ref{fig3}(a)]. Similar spreading characteristics are observed in the dynamics for displacement initial excitations as depicted in Fig.~\ref{fig3}(b).

For a more quantitative description, we follow the time evolution of the participation number $P$ and the second moment $m_2$, of the energy distribution to characterize the localization and spreading properties of the wave-packet. The former quantity $P$, is used to quantify the number of highly excited sites in a lattice~\cite{flachbook} and is computed as
\begin{equation} \label{eqp}
P = 1/ \sum h_n^2,
\end{equation}
where $h_n = H_n / H$ is the normalization of the site energy $H_n$. In the case of equipartition, for a lattice of size $N$, $ P = N$, while the other extreme gives $ P = 1$, for a wave-packet in which only a single site is highly excited.
The second moment $ m_2 $~\cite{Kundu} of the energy distribution given by
\begin{equation} \label{eqm2}
m_{2} =  \sum_n (n-\overline{n})^2 H_n /H,
\end{equation}
is a measure of the wave-packet's extent where $\overline{n} = \sum_n n H_n/H$ is the mean position of the energy distribution.

The time evolution of $\langle P \rangle$ indicates that a  maximum value is reached and remains constant for both velocity and displacement excitations as illustrated in Fig.~\ref{fig3}(c) by the black and magenta curves respectively. In this work, $ \langle \cdot \rangle$ denotes averages over $100$ disorder realizations. One of the differences between the two cases is that velocity initial excitations yield slightly higher values of $\langle P \rangle$ when compared to the $\langle P \rangle$ reached for displacement initial excitations. This is due to the fact that more low frequency propagating modes are excited for the case of velocity initial excitation than with displacement initial excitation \cite{Kundu}. 
The time evolution of $\langle m_ 2 \rangle$ for the two excitations is also different for the same reason.
 
 Regarding the second moment $m_2$, the usual practice is to assume that $\langle m_2 \rangle  \propto T^{\beta}$. Then the parameter $\beta$ is numerically estimated by the time local derivative
\begin{equation} \label{eqb}
\beta=  \frac{ d  \log_{10}\langle m_2 (T) \rangle }{ d  \log_{10} T}.
\end{equation}
The exponent $\beta$ is used to quantify the asymptotic behavior of $\langle m_2 \rangle$ for sufficiently large times. It is calculated by first smoothing the $m_2 (T)$ values of each disorder realization through a locally weighted difference algorithm~\cite{cleveland1,cleveland2} and then averaged over all realizations. The computed $\beta$ for the two cases, saturates to $\beta = 1.5$ and $\beta = 0.5$, for respectively velocity and displacement single site excitations as shown in Fig.~\ref{fig3}(d). This is an expected result because we have already shown the system to be practically a linear disordered FPUT lattice, since the translational DoFs do not couple to the rotational DoFs. 
Here we note that, according to the full system of equations [Eqs.~(\ref{em1})-(\ref{em2})] single site translational excitations as given by Eq.~(\ref{eq7}) will never couple the two DoFs. This is a particularity of the structural geometry under consideration. Thus,  whatever the initial excitation energy, single site initial translations will always lead to the linear behavior summarised in Fig.~\ref{fig3}. For this reason we shall not consider such initial conditions in Section~\ref{sec3}. 

\subsubsection{ Rotational degrees of freedom} \label{sssecR}

\begin{figure} 
\centering
\includegraphics[width=6.20cm]{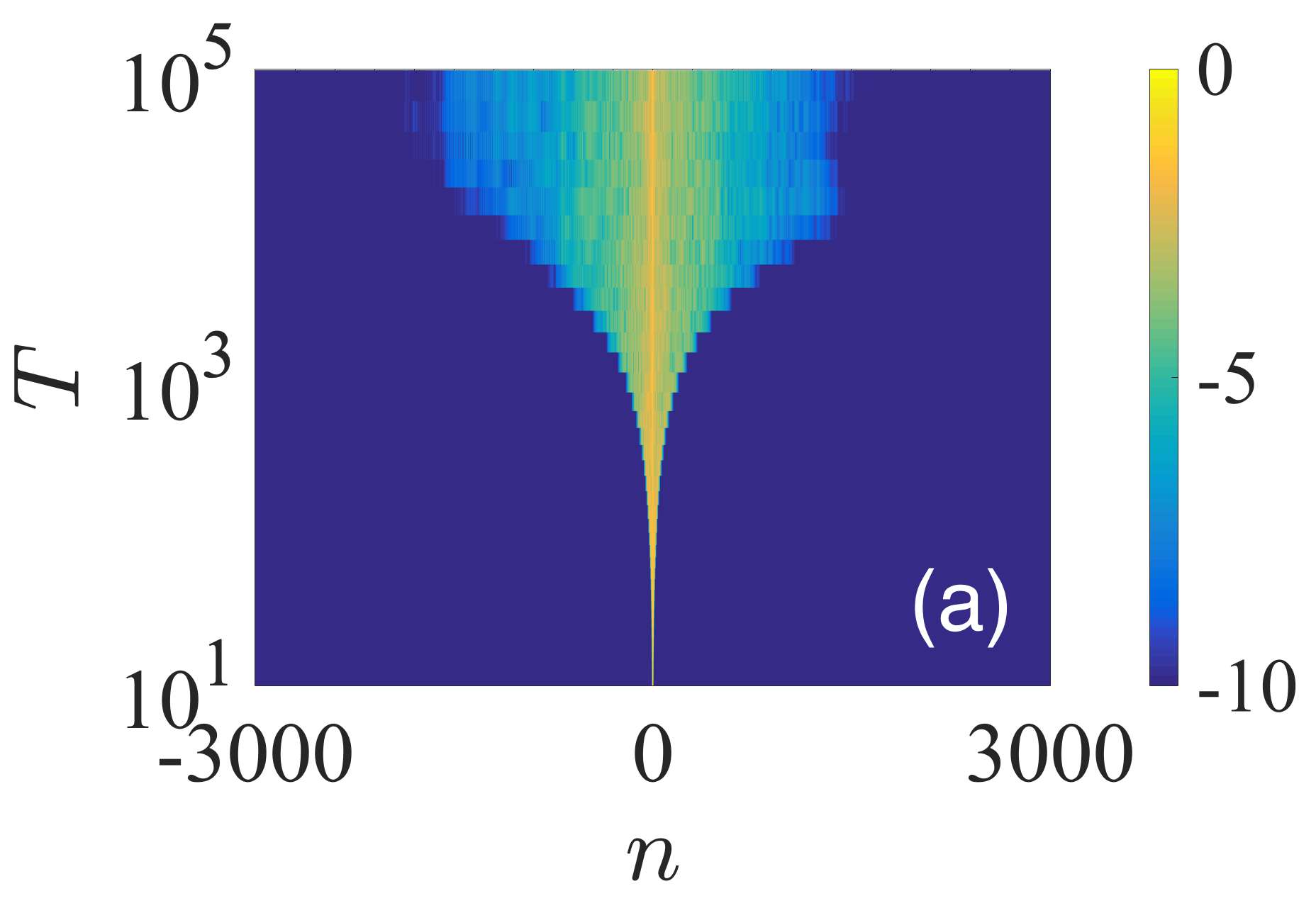}
\includegraphics[width=6.20cm]{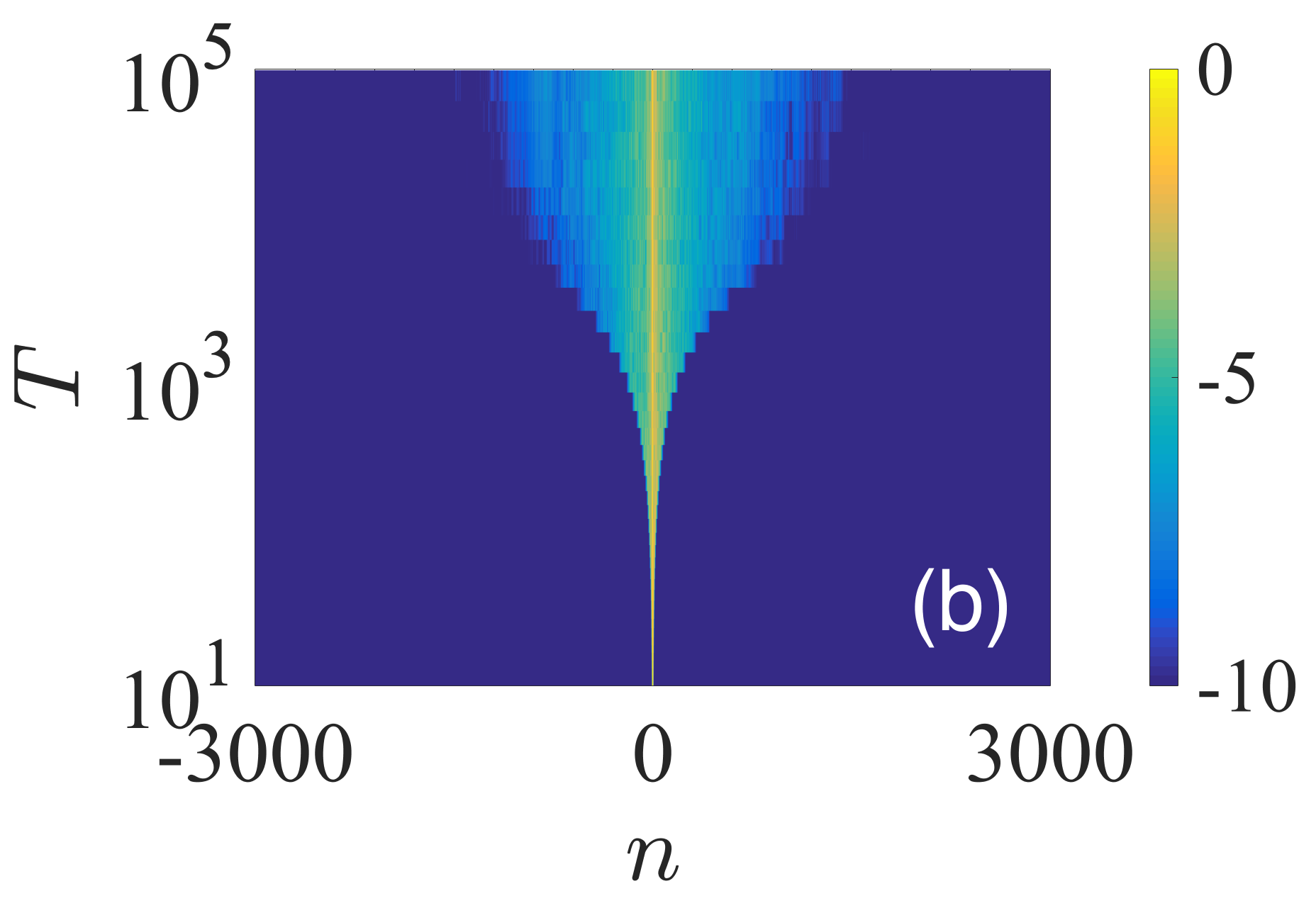}\\
\includegraphics[width=6.20cm]{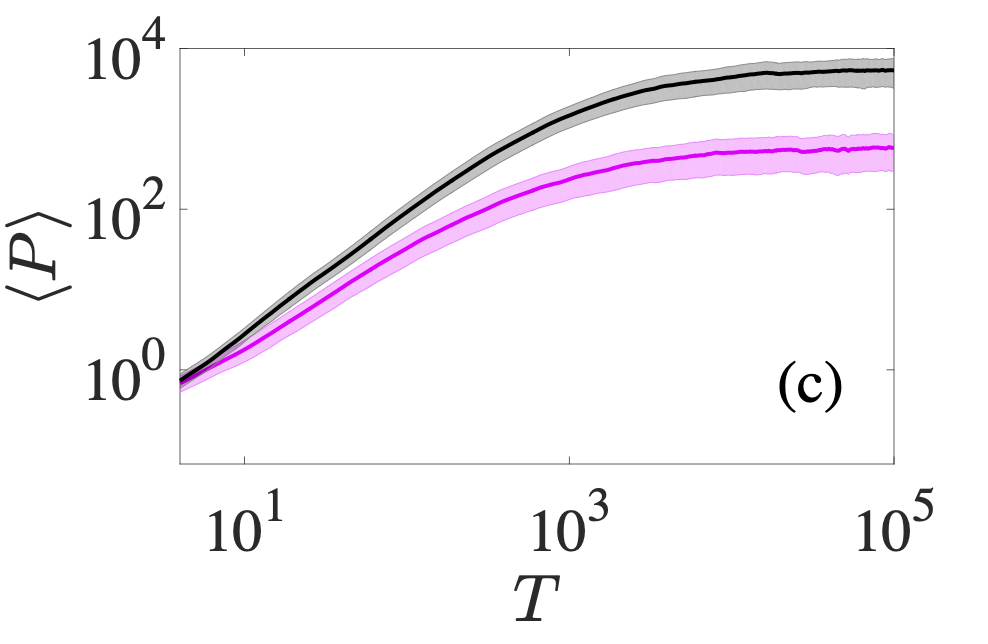}
\includegraphics[width=6.20cm]{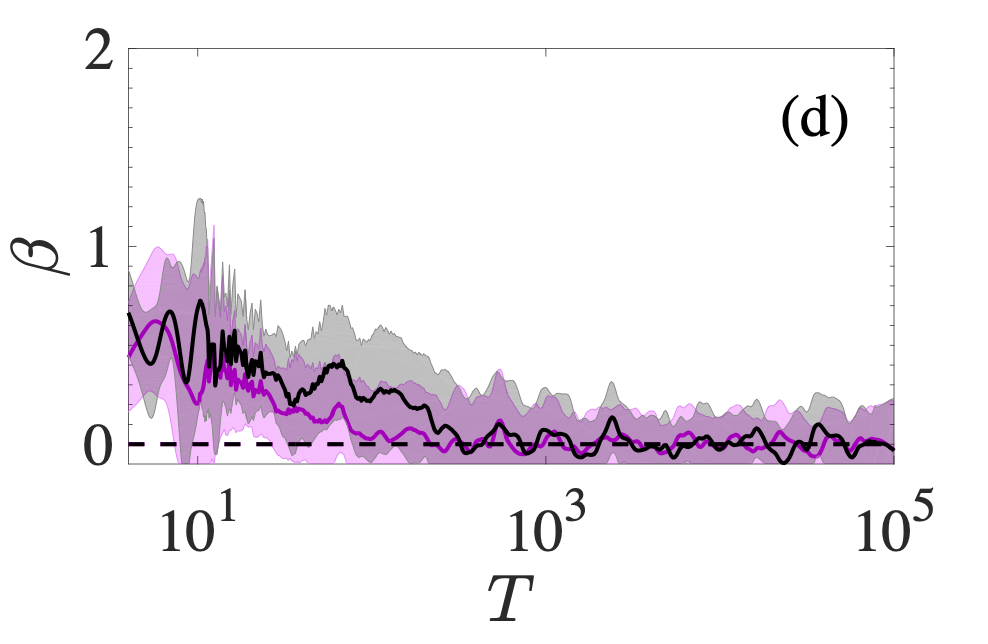}
\caption{\label{fig4} Similar to Fig.~\ref{fig3} but for initial angular deflections (magenta curves) and angular deflection time derivatives (black curves). The vertical axes for (a) and (b) are given in $\log_{10} $. The dashed line in (d) indicates $\beta=0$. }
\end{figure}

We now turn our attention to the dynamics of single site rotational excitations (angular deflection or the angular deflection time derivative). These excitations are expected to follow the dynamics of a discrete linear KG lattice as already explained earlier in Section.~\ref{ssec1}. More specifically, we consider independently the initial excitations
\begin{equation}
\dot{\theta}_{N/2}(0)=\sqrt{2H/\Gamma_{N/2}},\quad \textrm{or} \quad \theta_{N/2}(0)= \mu,
\label{eq72} 
\end{equation}
where $\mu$ is real and is chosen to match the desired system energy $H=H_0$.

The energy distribution profiles for both angular deflection time derivative (black curves) and angular deflection (magenta curves) excitations [see Figs.~\ref{fig4}(a) and (b)] show localized energy distributions for times $T \gtrsim 10^4$ after an initial phase of wave-packet spreading.
The subtle differences between angular deflection time derivative and angular deflection initial excitations in panels (a)-(b) of Fig.~\ref{fig4} are mainly due to the differences in the initially excited modes of the system for each respective excitation. Thus, the dynamics of the rotations in the linear limit show complete Localization in sharp contrast to the perpetual wave-packet spreading observed for the translational DoFs [Figs.~\ref{fig3}(a) and (b)]. This is a consequence of the fact that the KG system Eq~(\ref{em4}) describing rotations, is known to map to the DNLS and experiences localization of all modes \cite{kivshar,kivshar1993,johansson2006}. To quantify the localization for the initial angular deflections (magenta curves) and angular deflection time derivatives (black curves) we plot the time evolution of $ \langle P \rangle $ which reaches constant finite values after $ T \gtrsim 4 \times 10^{4} $ [Fig.~\ref{fig4}(c). A similar behavior is observed for $ \langle m_2 \rangle $ (not shown here) for  $ T \gtrsim 10^{3} $ ] in agreement to what is expected for a linear disordered KG chain. The final saturation of the $ \langle m_2 \rangle $ value is clearly depicted in the evolution of the exponent $\beta$ from the relation $ \langle m_2 \rangle \propto T^{\beta} $, which eventually becomes $ \beta =0 $ as clearly seen in Fig.~\ref{fig4}(d), showing that indeed the system is effectively a $1$D linear disordered KG chain.

\section{ Disordered nonlinear system} \label{sec3} 

Having considered the behavior of the system in the linear limit, we further study the fully nonlinear system as described by Eqs.~(\ref{em1})-(\ref{em2}). Before proceeding further, we reiterate that single site initial translational velocity ($\dot{U}_{N/2}$) or displacement ($U_{N/2}$) excitations \textit{do not induce a nonlinear response}. However, initial rotational excitations induce both a nonlinear response on rotations as well as nonlinear coupling between the two sets of DoFs. For the rest of this work we focus exclusively on single site (in the centre of the lattice) initial conditions of angular deflections as well as angular deflection time derivatives. 

\subsection{Weakly nonlinear regime} 
\begin{figure}
\begin{centering}
\includegraphics[width=14.0cm]{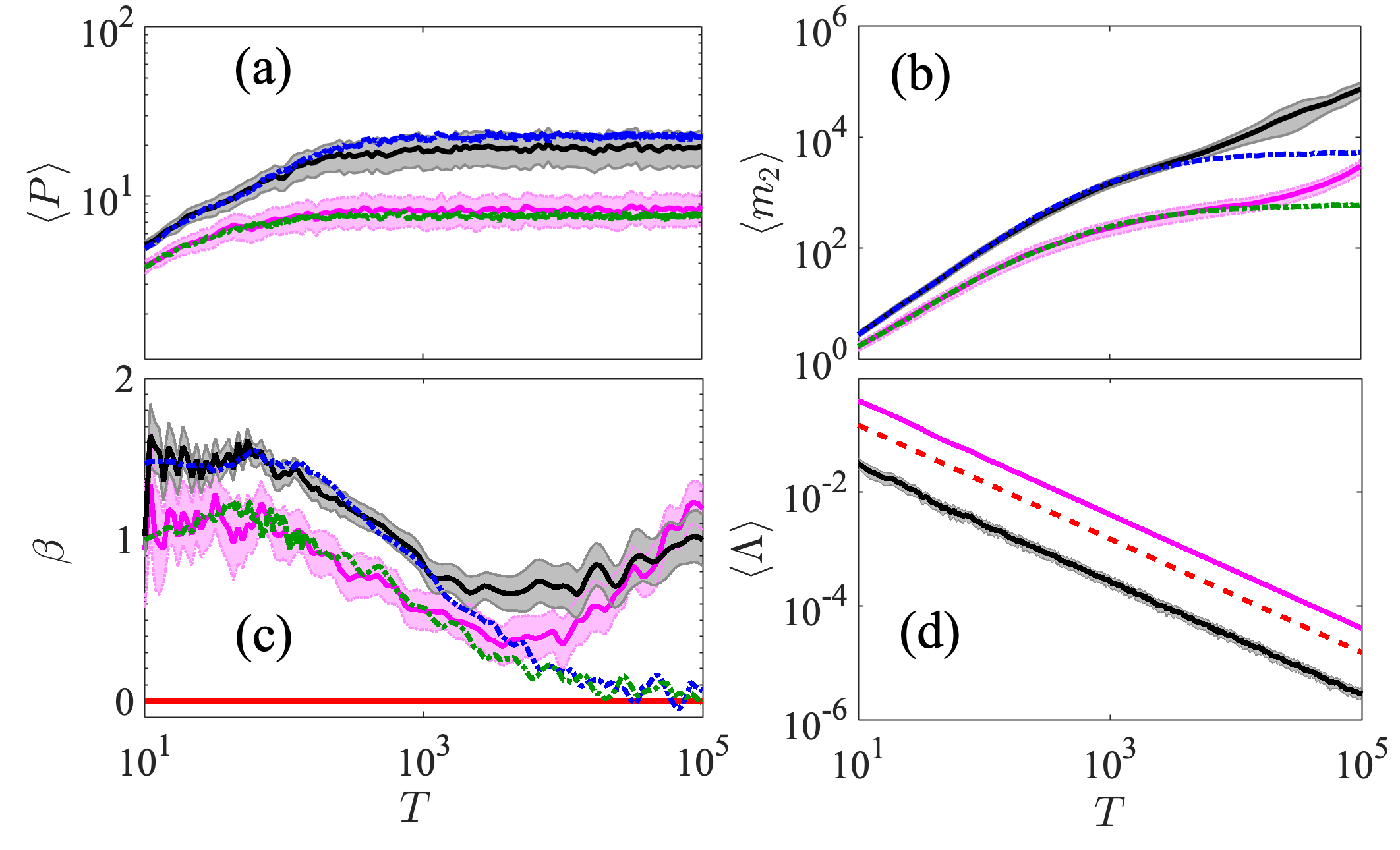}
\end{centering}
\caption{\label{fig5} Weakly nonlinear system,~ (a)-(d) Time evolution of $ \langle P \rangle $ (\ref{eqp}), $\langle m_2 \rangle$ (\ref{eqm2}), $\beta$ (\ref{eqb}) and $\langle \Lambda \rangle$ (\ref{eq41}), respectively. The black (magenta) curves correspond to single site initial angular deflection time derivatives (angular deflection) excitations. The average values $\langle \cdot \rangle$ are computed from $100$ disorder realizations and the lightly shaded areas represent the statistical error (one standard deviation). The dashed blue and green curves which are not always visible due to their overlapping with other curves, respectively show results of the linearized system for the initial conditions given by Eq.~(\ref{eq72}). All results are for the weakly nonlinear regime with $H = 10^{-8}$. The horizontal line in (c) indicates $\beta =0$. }
\end{figure}

We start by implementing single site angular deflections and time derivatives of angular deflections as initial excitations for a sufficiently small energy ($ H = 10^{-8}$) so that the system is in the weakly nonlinear regime. With the physical system at hand, this energy corresponds to an initial angle deflections of $\approx 0.2^{\circ}$. The time evolution of $\langle P \rangle$ is shown in Fig.~\ref{fig5}(a) where we observe a saturation to a constant value for each type of excitation [angular deflection time derivative (black curve) and angular deflection (magenta curve)]. In fact we also plot, in the same figure, the corresponding linear result (dashed curves) and we observe that the weak nonlinearity does not affect the participation number. Nevertheless, we find that nonlinearity plays a significant role on the evolution of $\langle m_2 \rangle$ and $\beta$. This is illustrated in Figs.~\ref{fig5}(b)-(c) where both these quantities are found to increase for the last two decades ($T \gtrsim 10^3$) of the evolution clearly indicating energy spreading  in the system.

To further explore the spreading it is worthwhile to also investigate the chaoticity of the system using the finite time maximum Lyapunov exponent (ftMLE). It is often found, in systems with multiple degrees of freedom, that energy spreading is due to chaos around the excitation region \cite{disorder_15,skokos2013,Bob2018}. The ftMLE,
\begin{equation} \label{eq41}
\Lambda(T) = \frac{1}{T} \ln \frac { || \mathbf{w} (T) || }{ || \mathbf{w} (0) ||},
\end{equation}
is computed using the so-called standard method \cite{benettin,lec_notes}. $\mathbf{w} (T)$ is a vector of small perturbations from the phase space trajectory at time $T$ (also called deviation vector) which we denote as 

\begin{equation}
\begin{aligned}
 \mathbf{w} (T)  = & [ \delta  U_1 (T), \ldots , \delta  U_N (T), \\ & \delta  \theta_1 (T), \ldots , \delta  \theta_N (T), \\
 &M_1 \delta \dot{U}_1 (T),  \ldots , M_N \delta \dot{U}_N (T), \\
 &\Gamma_1 \delta  \dot{\theta}_1 (T),  \ldots , \Gamma_N \delta \dot{\theta}_N (T)], 
 \end{aligned}
\end{equation}
where $\delta  U_n (T)$ and $\delta  \theta_n (T)$ indicates small perturbations in positions, while $M_n \delta \dot{U}_n (T)$ and $\Gamma_n \delta \dot{\theta}_n (T)$ indicate small perturbations in momenta for the two DoFs at the $n$th lattice site. The mLE is defined as $ \lambda = \lim_{T \to \infty} \Lambda (T) $. In Eq.~(\ref{eq41}), $|| \cdot ||$ denotes the usual Euclidean vector norm. For chaotic trajectories, $\Lambda $ attains a finite positive value, otherwise, $\Lambda \propto T^{-1}$ for regular orbits. An efficient and accurate method to follow the evolution of $\mathbf{w} (T)$ is to numerically integrate the so-called variational equations \cite{galgani}, which govern the vector's dynamics, together with the Hamilton equations of motion using the tangent map method outlined in Refs.~\cite{gerlach1,gerlach2,gerlach3}. The mLE can be used to discriminate between regular and chaotic motions since $\Lambda  = 0$ for regular orbits and $\Lambda > 0$ for chaotic orbits. The magnitude of the mLE can also be used as a measure of the chaoticity: larger mLE values imply stronger chaotic behaviors.

 In Fig.~\ref{fig5}(d) we show the calculated ftmLE which is found to follow the power law decay $ \langle \Lambda \rangle \propto T^{-1}$,  and thus the system exhibits regular dynamics. As such we conclude that  the observed energy spreading \textit{cannot} be attributed to chaoticity as is the case for other lattice models including the single DoF per site KG lattice \cite{skokos2013,Bob2018}.

\begin{figure}
\centering
\includegraphics[width=12.0cm]{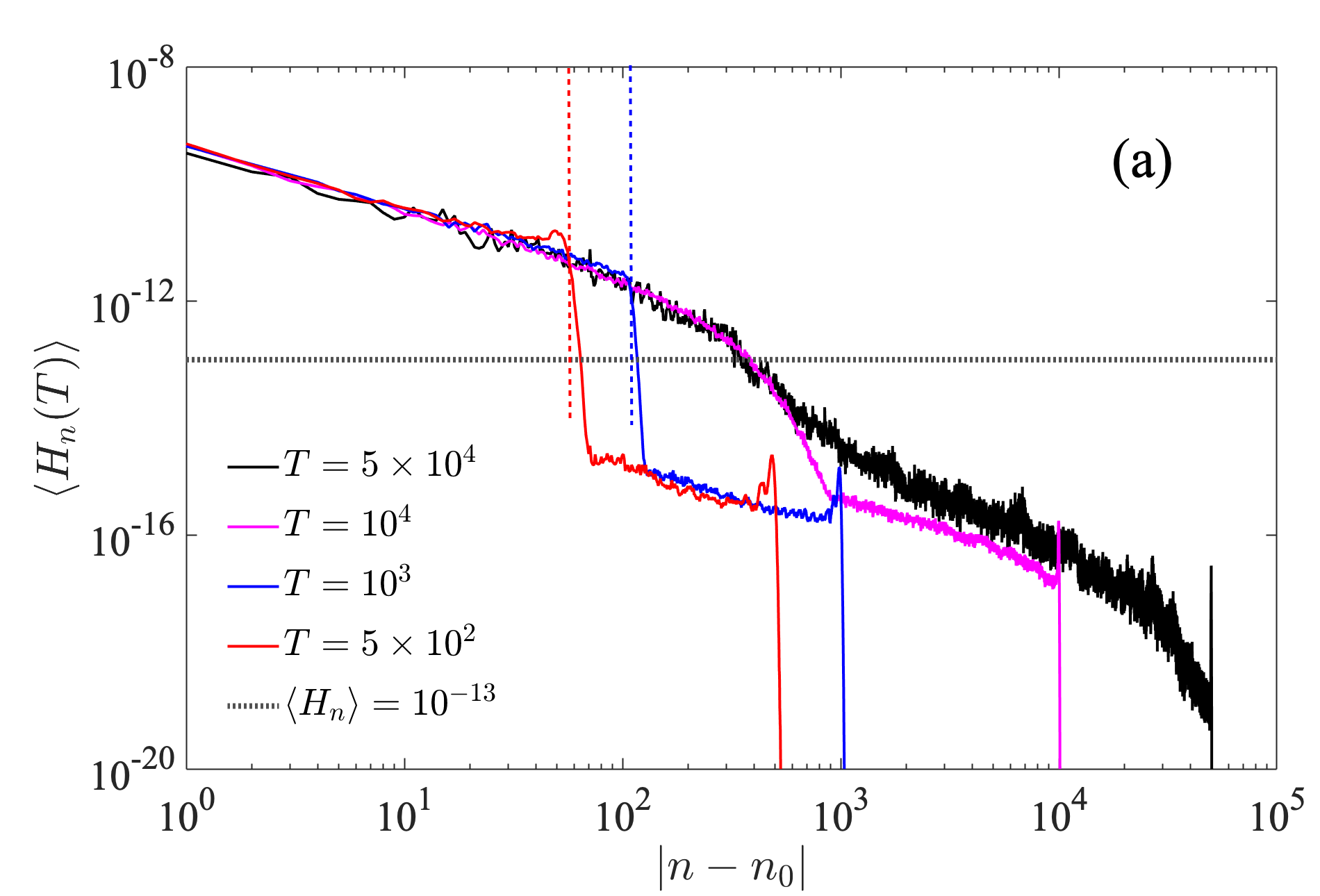}\\
\includegraphics[width=12.0cm]{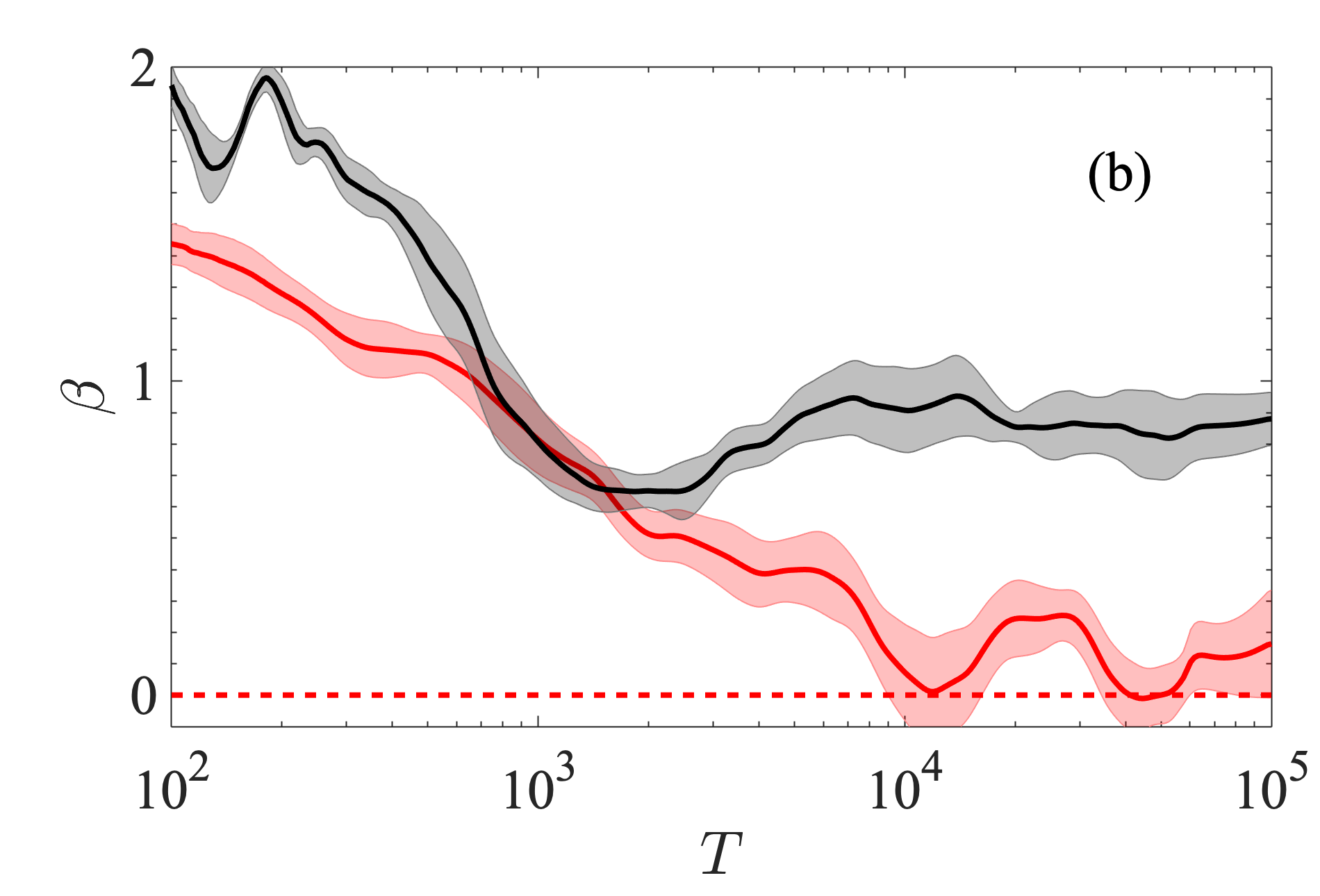}
\caption{\label{fig6} (a) Average energy distribution profiles of $100$ disorder realizations for $T= 5\times 10^2$~(red curve) , $T= 10^3$~(blue curve), $T= 10^4$~(magenta curve) and $T= 5\times 10^4$~(black curve). The dotted horizontal line marks the energy $ \langle H_n \rangle = 10^{-13}$ while the vertical dashed lines indicate the position of the slower wave-front at the indicated times. (b) Time evolution of the exponents $\beta$ for the central part (solid red curve) and the tails (solid black curve). The lightly shaded regions in (b) indicate the statistical error (one standard deviation). All panels are for angular deflections of system energy $H=10^{-8}$.}
\end{figure}

In order to explain the spreading we need to further monitor the energy density in the lattice as a function of time. In Fig.~\ref{fig6}(a) we show four snapshots of the energy density around the initial excitation point ($n=0$). The spatial distribution of energy is separated into two distinct regions: a large amount of energy localized around  $n=0$  and an extended tail with much lower energy (five orders of magnitude less). In fact, if we consider the propagation of the higher energy central part, especially for $T \lesssim 10^3$, we clearly observe a leading wave-front (dashed vertical lines) which propagates slower than the main leading wave-front. The latter corresponds to the low energy regions [$H_n \lesssim 10^{-13}$ in Fig.~\ref{fig6}(a)] which propagates with a normalized velocity close to one. On the contrary the high energy part around the center [$H_n \gtrsim 10^{-13}$] is propagating much slower with a normalized velocity of $\approx 0.1$. These two velocities correspond respectively to the largest group velocities of the translation and rotational branches of the dispersion relations shown in Fig.~\ref{fig2}. Thus we conjecture that the two distinct parts of the energy distribution, high and low, correspond respectively to the two different types of DoFs i.e., the rotational and translation deflections.

Furthermore, as shown by the snapshots for $T\gtrsim 10^4$ in Fig.~\ref{fig6}(a), at later times only the lower (translation) part of the energy continues to spread.
This fact is corroborated by calculating the exponent of the second moment for energies lower (larger) than a threshold ($H=10^{-13}$) as shown in Fig.\ref{fig6}(b). The exponent $\beta$ for the high energy central part almost vanishes (red curve), indicating no spreading while for the low energy region value of $\beta$ (black curve) is finite revealing spreading. Thus we conclude this sub-section by explaining the dynamics in the following manner. In the weakly nonlinear regime, the role of the nonlinearity is to induce spreading by stimulating the translation DoFs (which have an FPUT character) through the nonlinear coupling. In this way, the energy spreading in this regime is characterized by a hybrid of KG-like and FPUT-like behaviors. In addition the rate of wave-packet spreading as quantified by $\beta$ shows no defined asymptotic behavior but rather a dependency on time [Fig. \ref{fig5}(c)]. This is in accordance with the results obtained for a weakly nonlinear FPUT lattice by Lepri \textit{et.al.}, \cite{Lepri}.
 
\subsection{Strongly nonlinear regime} 
\begin{figure}
\begin{center}
\includegraphics[width=12.60cm]{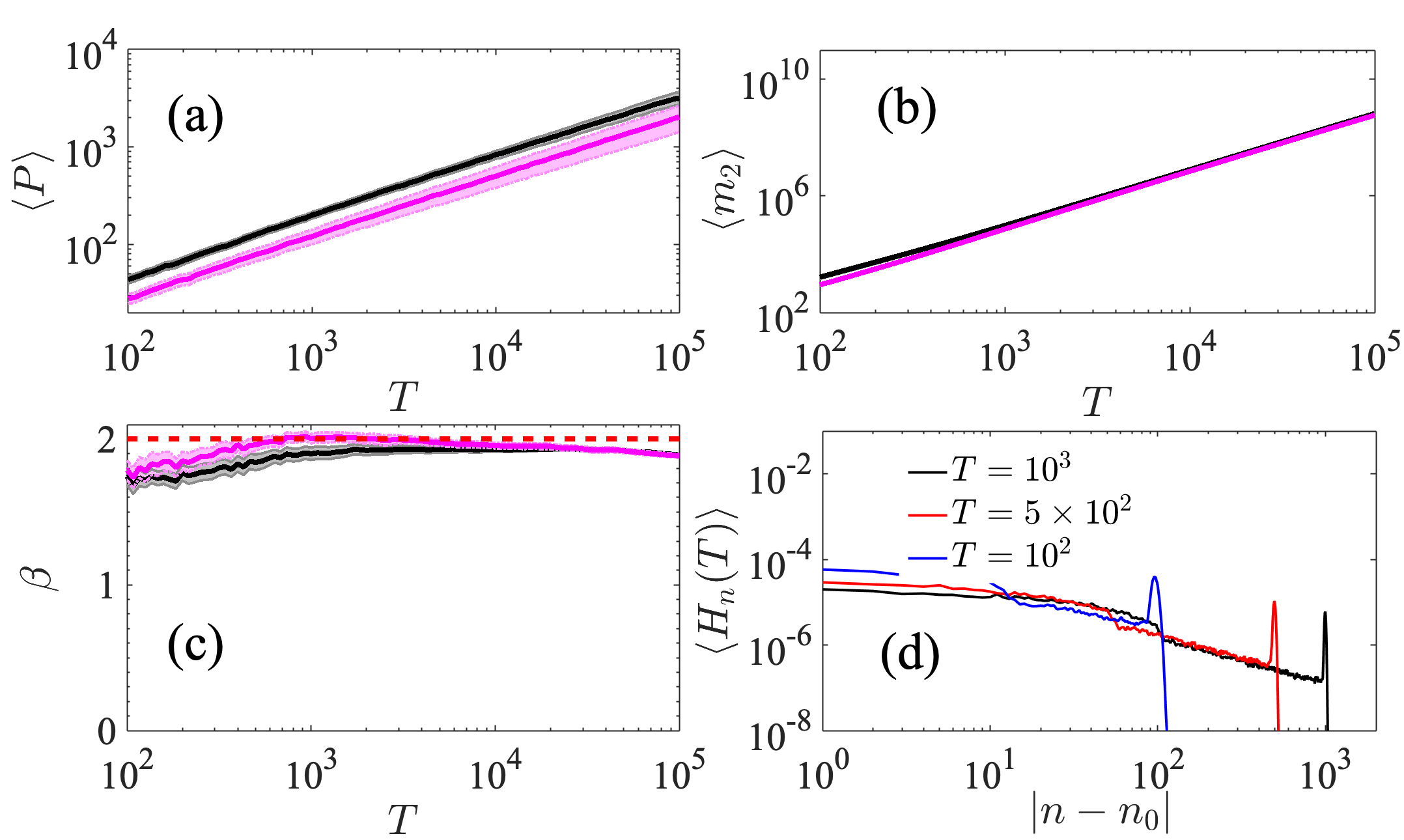}
\end{center}
\caption{\label{fig7} (a)-(c) Similar to Figs.\ref{fig5}(a)-(c). (d) Average energy density profiles over $100$ disorder realizations at $T=10^2$ (blue curve), $T=5 \times 10^2$ (red curve) and $T=10^3$ (black curve). Results in all panels are for energy $H= 5 \times 10^{-3}$.}
\end{figure}

Let us move away from the weakly nonlinear regime and increase the nonlinearity of the system by increasing the initial excitation energy to $ H = 5 \times 10^{-3}$. This energy corresponds to large initial angle deflections of about $30^{\circ}$. Note that this value of energy leads to strong nonlinear behavior. The system, in this regime, shows a completely different behavior of spreading as is indicated by the increase of $\langle P \rangle$ during the time evolution as shown in Fig.~\ref{fig7}(a). The number of highly excited sites grows in time contrary to what we observed for the weakly nonlinear regime, which shows practically no growth in $\langle P \rangle$ at long times. As expected, this increase of participating particles leads to wave-packet spreading and this is confirmed by the time evolution of $\langle m_2 \rangle$ which is also increasing as indicated in Fig.~\ref{fig7}(b). A feature we observe in this strong nonlinear regime is that the dynamics no longer depends on the type of initial condition (angular deflections or time derivatives of angular deflections). This is noticeable from the practically overlapping black (time derivatives of angular deflections) and magenta (angular deflections) curves in Figs.~\ref{fig7}(b)-(c). To quantify wave-packet spreading, we estimate the exponent of $\langle m_2 \rangle \propto T^{\beta}$ which is found to acquire values around $\beta \approx 2$ as shown in Fig.~\ref{fig7}(c). This value indicates a very strong spreading corresponding to a near ballistic propagation. We find this result to be quite interesting since the nonlinearity of the flexible architected material under study, is strong enough to bring the system to ballistic behavior, which is not always the case in other systems such as the FPUT and KG lattices~\cite{pikovsky,flachbook}. In fact, the other example to our knowledge, where near ballistic behavior in a disordered system is observed, is for mechanical lattices featuring non-smooth nonlinearities due to Hertzian forces \cite{jk}. 
 
 Also, in this strongly nonlinear regime the distinct behavior of the two types of DoFs that was observed in Fig.~\ref{fig6}(a) is now lost. According to Fig.~\ref{fig7}(d) showing the mean profile of the energy distribution, we identify a large part of the energy being localized around the center and a propagating tail  travelling almost ballistically. This is expected since according to Eqs.~(\ref{em1})-(\ref{em2}) the coupling of the two types of DoFs is enhanced at each lattice site when the rotations are of high amplitude. Thus we no longer distinguish between a KG-like and FPUT-like evolution of the energy profiles. We have also considered a range of system energies between $H=10^{-8} - 10^{-3}$ and found that the distinction between KG- and FPUT-like behaviors gradually disappears as the system energy is increased. Some of the results for the intermediate energies  are reported in Ref.~\cite{ngapasarephd}.
 
\begin{figure}
\begin{center}
\includegraphics[width=12.5cm]{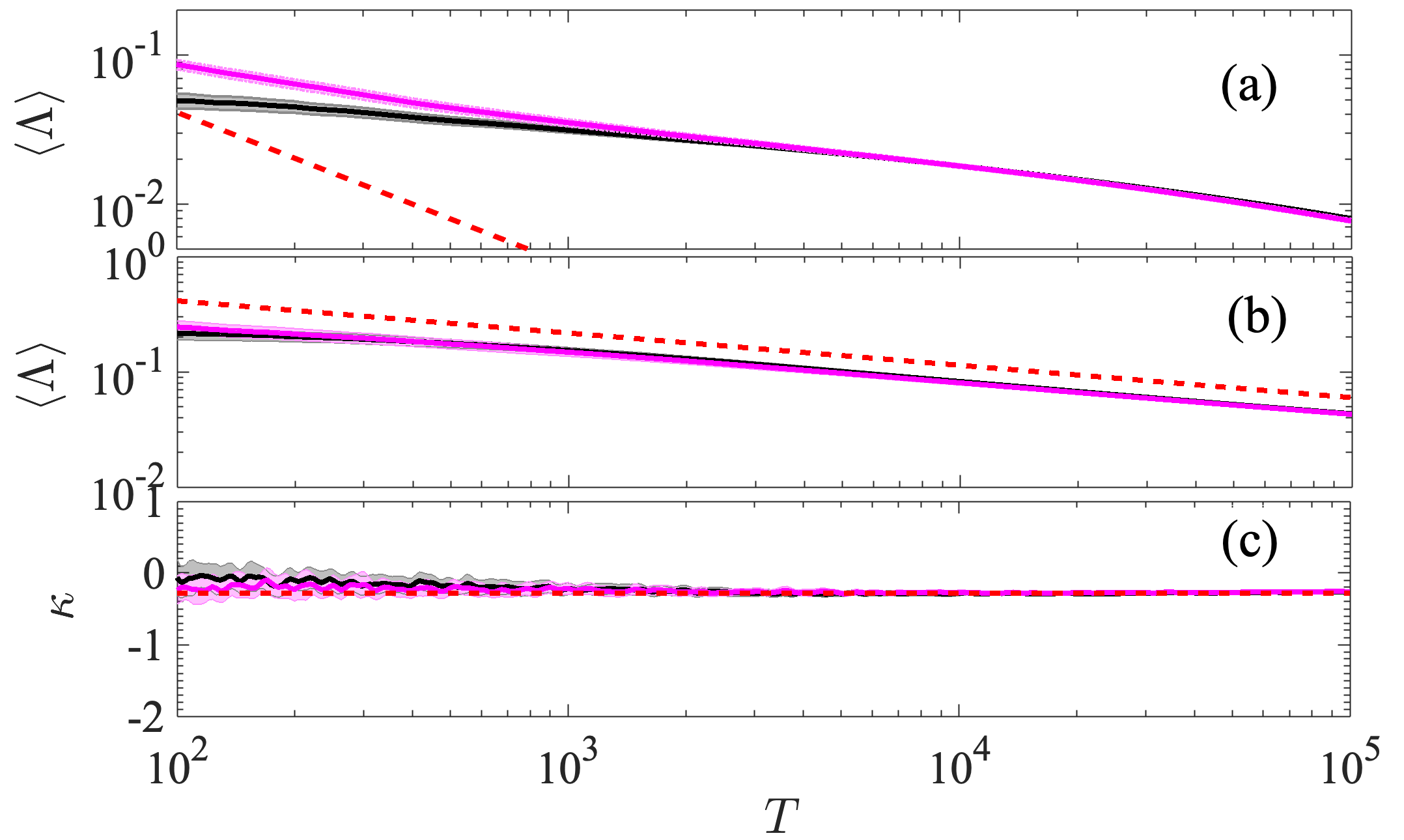}
\end{center}
\caption{\label{fig8}~(a) Time evolution of the average ftMLE, Eq.~(\ref{eq41}), $ \langle \Lambda \rangle $ for the strongly nonlinear regime angular deflections and time derivatives of angular deflections as initial excitations for energy $H=5 \times 10^{-3}$. The red dashed line indicates the power law $\langle \Lambda \rangle \propto T^{-1}$. (b) Similar to (a) but for energy $H=10^{-1}$. Here the red dashed line indicates the power law $\langle \Lambda \rangle \propto T^{\kappa}$, with $\kappa = -0.28 $. (c) Time evolution of the exponent $ \kappa $ at energy $H = 10^{-1}$. In all panels the magenta and black curves respectively show results for angular deflections and time derivatives of angular deflections as initial excitations. The magenta horizontal dashed line indicates $\kappa = -0.28$. }
\end{figure}

Regarding the chaoticity of the system, for such high initial angles and thus strong nonlinearity, we find the dynamics to be chaotic. This is evident in Fig.~\ref{fig8}(a) where the mean value of the ftMLE $\Lambda$ Eq.~(\ref{eq41}), is shown to be decreasing in a much slower rate compared to regular dynamics (dashed line). This type of chaotic behavior, where the ftMLE does not reach an asymptotic constant value, has recently attracted much attention and appears to be a particular case of chaos spreading \cite{disorder_15, Bob2018, many2} and it is related to the fact that as the wave-packet spreads, the constant total energy is shared among more sites and consequently the energy per excited site (which plays the role of active nonlinearity strength) decreases. Consequently, the ftMLE which is a global measure of chaos, is also decreasing. 

However, different to what was found in Refs.~\cite{disorder_15, Bob2018, many2}, for the architected lattice under study, here the slope of $ \langle \Lambda \rangle$ does not reach a constant value slope even for the \textit{largest possible } angular deflection  of $45^{\circ}$.
Thus, for the sake of completeness and to be able to compare the results regarding the chaos spreading of the soft architected lattice with other models in the literature, we extend our numerical simulations using even larger initial energies. In particular, in Figs.~\ref{fig8}(b)-(c) we show results using an initial energy of $H=10^{-1}$.
Note that this value of energy leads to an even stronger nonlinear behavior than the penultimate case above. 
%
 %
 %
 %
%
We also observe in Fig.~\ref{fig8}(b) that the ftMLE reaches a clearly constant slope with which it is decaying. In particular, the slope of the exponent assuming that $\langle \Lambda \rangle \propto T^{\kappa}$  is found to be approximately $\kappa = -0.28$ for both angular deflections and time derivatives of angular deflections. Note that this asymptotic value of the exponent $\kappa$, is comparable to the corresponding values for the ftMLE in the more studied cases of $1$D lattices namely the disordered DNLS equation and the disordered KG model \cite{skokos2013, Bob2018}. More precisely, the value  $-0.28$ lies between the values obtained for the so-called weak ($\kappa = - 0.25$) and strong chaos ($\kappa = - 0.3$) regimes of these $1$D models having a single DoF per lattice site. 

\section{Summary and conclusions} \label{sec4} 

We have studied numerically energy spreading and chaos in a nonlinear disordered architected mechanical lattice. The lattice under consideration describes rotating LEGO bricks connected with flexible links which was recently studied experimentally. The in-plane motions of the lattice are described by two DoFs per lattice site i.e., translations and angular deflections. Furthermore, in the linear limit of the aligned structure, the two DoFs per site are completely decoupled. In this state, the lattice shows two distinct behaviors corresponding to the FPUT and KG-like behaviors for translational and rotational DoFs respectively. For both cases, we review results regarding the behavior of energy spreading  under the effect of disorder.

Using single site angular deflections and time derivatives of angular deflections initial excitations, we studied the system for different strengths of nonlinearity focusing on the weakly nonlinear and strongly nonlinear regimes. For the weakly nonlinear regime we have shown that the total energy density of the lattice is split into two parts: i) a slow spreading part around the excitation point following the KG-like behavior and ii) the fast propagating tails of lower energy which travel with the speed of sound of the corresponding FPUT lattice and are responsible for the evolution of the second moment of the total energy distribution. 
For sufficiently large initial excitations, the strong nonlinearity of the flexible architected lattice forces the initial wave-packet to spread ballistically and the distinction between a KG- and an FPUT-like behavior is lost. We note here that a ballistic behavior under strong disorder is not easily achieved and here the responsible  physical mechanism is the large geometrical nonlinearity. 

Additionaly, we show that chaos is found to persist during the energy spreading although its strength decreases in time as quantified by the evolution of the system's ftMLE. Here, the power law time evolution of the exponent of the ftMLE is found to acquire a value which lies between the ones obtained for the so-called weak and strong chaos regimes of the well studied nonlinear KG lattice.

Our results show that flexible architected elastic lattices with coupled DoFs per site provide a modern physical platform to study and observe rich wave dynamics which can not otherwise be seen with classical uncoupled fundamental models. Some interesting directions arise from our results like the study and manipulation of energy propagation by tuning the dispersion characteristics of rotations, by changing the shear and bending stiffness's. Furthermore, here we only considered an aligned structure where the two DoFs per site are uncoupled in the linear limit. Extensions to other geometries, where the linear modes are polarised will probably reveal a variety of spreading characteristics and provide a means of controlling energy transport in highly heterogeneous lattices. 

\section*{Acknowledgements}
A.N. acknowledges funding from the University of Cape Town (University Research Council, URC) postdoctoral Fellowship
grant and the Oppenheimer Memorial Trust (OMT) postdoctoral Fellowship grant. Ch. S. thanks the Universit\'{e} du Mans for its hospitality during his visits when part of this work was carried
out. We also thank the Centre for High Performance Computing \cite{chpc} for providing
computational resources for performing significant parts of this paper’s computations. We also thank V. Tournat for useful discussions.

\section*{Author Declarations}
The authors have no conflicts to disclose.

\section*{Data Availability Statement}
The data that support the findings of this study are available from the corresponding author upon reasonable request.

\bibliographystyle{unsrt}
\bibliography{aipsamp}

\end{document}